# Emergence of nontrivial magnetic excitations in a spin liquid state of kagomé volborthite

D. Watanabe[1,2], K. Sugii[1], M. Shimozawa[1], Y. Suzuki[1], T. Yajima[1], H. Ishikawa[1], Z. Hiroi[1], T. Shibauchi[3], Y. Matsuda[2], M. Yamashita[1*]

[1] Institute of Solid State Physics, the University of Tokyo, Chiba 277-8581, Japan.

[2] Department of Physics, Kyoto University, Kyoto 606-8502, Japan.

[3] Department of Advanced Materials Science, University of Tokyo, Chiba 277-8561, Japan.

*Correspondence to: my@issp.u-tokyo.ac.jp

**When quantum fluctuations destroy underlying long-range ordered states, novel quantum states emerge. Spin-liquid (SL) states of frustrated quantum antiferromagnets, in which highly-correlated spins keep to fluctuate down to very low temperatures, are prominent examples of such quantum states. SL states often exhibit exotic physical properties, but the precise nature of the elementary excitations behind such phenomena remains entirely elusive. Here we utilize thermal Hall measurements that can capture the unexplored property of the elementary excitations in SL states, and report on the observation of anomalous excitations that may unveil the unique features of the SL state. Our principal finding is a negative thermal Hall conductivity $\kappa_{xy}$ which the charge-neutral spin excitations in a gapless SL state of the two-dimensional kagomé insulator volborthite $Cu_3V_2O_7(OH)_2 \cdot 2H_2O$ exhibit, in much the same way in which charged electrons give rise to the conventional electric Hall effect. We find that $\kappa_{xy}$ is absent in the high-temperature paramagnetic state and develops upon entering the SL state in accordance with the growth of the short-range spin correlations, demonstrating that $\kappa_{xy}$ is a key signature of the elementary excitation formed in the SL state. These results suggest the emergence of nontrivial elementary excitations in the gapless SL state which feel the presence of fictitious magnetic flux, whose effective Lorentz force is found to be less than 1/100 of that experienced by free electrons.**

SLs are novel states which can occur in a magnetic system when the underlying magnetic order gives way to quantum fluctuations (1). In such states the constituent spins are highly correlated but continue to fluctuate strongly down to temperatures much lower than the spin interaction energy scale, $J$. Novel notions such as emergent gauge fields, topological order and fractionalized excitations have been associated with collective phenomena in SLs. In particular, both experiments (2-5) and theories (6-11) suggest that SL states display many unusual properties. It has been reported, for instance, that low-energy spin excitations in organic insulators with a triangular lattice structure behave like mobile carriers in a paramagnetic metal with a Fermi surface (2,3), in contrast to the charge degree of freedom which is gapped. A description in terms of a SL state with fractionalized spin excitations was incorporated to account for the excitation continuum signal detected in a kagomé antiferromagnet (4). A magnetization transport measurement has shown that a pyrochlore frustrated magnet exhibits the characteristics of a supercooled spin liquid state (5). Exotic quasiparticles such as spinons (6-8), visons (9, 10) and photons (11) have been predicted theoretically. Despite these intensive activities, the precise character of the elementary excitations in SL states remain, from an experimental point of view, to be pinned down.

In conducting systems, it is the charge transport properties that act as the window through which we accumulate information that are essential in unraveling the physics of novel electronic states such as the quantum Hall states and other non-Fermi-liquids. Likewise, in insulating quantum magnets, thermal-transport measurements have been proven to be a powerful probe in unveiling the ground state and quasiparticle excitations (2, 12-16). Recently, theoretical works have suggested that thermal Hall measurements provide new insights into the nature of exotic excitations in magnetic insulators (17-20). In conducting systems, the electrical Hall ($\sigma_{xy}$) and thermal Hall ($\kappa_{xy}$) conductivities are related by the Wiedemann-Franz law $\kappa_{xy}/T = L\sigma_{xy}$ (ref. 21), where $L$ is the Lorentz number. In magnetic insulators, in contrast, there are no charged currents, and thus a magnetic field cannot exert a Lorentz force. Nevertheless the thermal Hall effect has been predicted to occur in both ordered (17,18) and disordered magnets (17,19,20) as a result of the intrinsic Berry phase curvature and an emergent gauge field, respectively. Indeed, a finite $\kappa_{xy}$ was reported in the ferromagnetic ordered state of magnetic insulators (12, 13). Very recently, its observation was also reported in the disordered states of the spin ice compound Tb$_2$Ti$_2$O$_7$ (ref. 14) and ferromagnetic kagomé Cu(1,3-bdc) (ref. 15). We note, however, that in the former study, a finite $\kappa_{xy}$ was observed in the paramagnetic phase far above the temperature corresponding to $J$ ($T \gg J/k_B \sim 1$ K) (ref. 22). Meanwhile, in the latter experiment there is no SL phase owing to the absence of geometrical frustration. In yet another study, thermal Hall measurements were performed in the SL state of triangular organic compound EtMe$_3$Sb[Pd(dmit)$_2$]$_2$, but no discernible $\kappa_{xy}$ signal was observed (2). Thus the experimental verification of thermal Hall conductivity in SL states remains to be a subject of vital importance.

## Results

**Heat capacity and magnetic susceptibility measurements.** Volborthite, Cu$_3$V$_2$O$_7$(OH)$_2$·2H$_2$O, is a magnetic insulator, in which two dimensional (2D) layers of Cu$^{2+}$ have a distorted kagomé structure with inequivalent exchange interactions(23, 24, 25, 26) (see the inset of Fig. 1). The temperature dependence of the magnetic susceptibility $\chi$ shows a behavior that is typical of 2D frustrated spin systems (Fig. 1). Below $T^* \sim 60$ K, $\chi(T)$ begins to deviate from the paramagnetic Curie-Weiss behavior (24, 27, 28), implying that the spin correlations grow gradually when the temperature energy scale $k_B T^*$ becomes comparable to the effective spin interaction energy $J_{eff}$. A peak of $\chi(T)$ at $T_p \sim 18$ K suggests that spins are highly correlated at lower temperatures. Recently, based on the observation of an extremely wide one-third magnetization plateau (26), effective spin Hamiltonians featuring coupled frustrated chains (27) and a coupled trimers (28) were proposed. In the latter framework, the system can be viewed as a frustrated triangular lattice with competing antiferromagnetic and ferromagnetic interactions with exchange couplings whose energy scales are not far from $k_B T^*$. In either models, the combination of strong geometrical frustration with enhanced quantum fluctuations for $S = 1/2$ suppresses the magnetic ordering down to $T_N \sim 1$ K ($\sim$ 2 K) at zero field (15 T) (ref. 24, 25), which infers the presence of a SL state in a wide temperature range $T_N < T < T^* \sim J_{eff}/k_B$. Clearly the $\chi$ above $T_N$ extrapolated to $T = 0$ remains finite, suggesting the gapless nature of the spin excitations in the SL state. A strong additional evidence supporting the presence of the gapless excitations comes from the specific heat measurements at low temperatures (Fig. 1), which show a large linear temperature-dependent contribution, $C/T$ ($T \to 0$) $\sim 50$ mJ K$^{-2}$ mol-Cu$^{-1}$.

**Longitudinal and transverse thermal conductivities.** Figure 2 shows the temperature dependence of the longitudinal thermal conductivity ($\kappa_{xx}/T$) in zero field and in a magnetic field of 15 T (H⊥2D plane). In the present system, heat is transferred by spin excitations and phonons: $\kappa_{xx} = \kappa_{xx}^{\text{spin}} + \kappa_{xx}^{\text{ph}}$. Below around $T_p$, $\kappa_{xx}/T$ is suppressed by the magnetic field. We point out that this suppression arises from $\kappa_{xx}^{\text{spin}}$ because of the following reasons. The field dependence of $\kappa_{xx}^{\text{ph}}$ is determined by spin-phonon scattering, which contains elastic and inelastic processes. The elastic scattering process is suppressed by the alignment of spins with the magnetic field. The inelastic scattering is directly related to the quantum dynamics of spin, which is also suppressed with field by the formation of Zeeman gap. Therefore an application of magnetic fields leads to an enhancement of $\kappa_{xx}^{\text{ph}}$.

The thermal conductivity of spin excitations can be expressed as $\kappa_{xx}^{\text{spin}} \sim C_{\text{spin}} v_{\text{spin}} \ell_{\text{spin}}$, where $C_{\text{spin}}$ is the heat capacity, $v_{\text{spin}}$ is the velocity and $\ell_{\text{spin}}$ is the mean free path of the elementary spin excitations. Since the magnetic field enhances $\ell_{\text{spin}}$ by aligning spins and $C_{\text{spin}}$ shows only a weak field dependence around 8 K (Fig. 1), the field suppression of $\kappa_{xx}^{\text{spin}}$ is dominated by the suppression of $v_{\text{spin}}$. Similar results have been reported in spin chain compounds (16) where the velocity of elementary excitations is suppressed by fields. The lower limit of $\ell_{\text{spin}}$ is simply estimated by assuming that $\Delta \kappa = \kappa_{xx}(H = 0) - \kappa_{xx}(H = 15 \text{ T})$ gives a lower limit of $\kappa_{xx}^{\text{spin}}$. From the assumption of a linear energy dispersion, the velocity is obtained as $v_{\text{spin}} \sim J_{eff} a/\hbar \sim 2.3 \times 10^3$ m/s, where $J_{eff} \sim 60$ K and $a \sim 2.9$ Å is the mean distance between the nearest Cu ions. From $C_{\text{spin}}/T \sim 50$ mJ K$^{-2}$ mol-Cu$^{-1}$ (Fig. 1) and $\Delta\kappa/T \sim 0.03$ W K$^{-2}$ m$^{-1}$ (Fig. 2) at 8 K, we find the lower limit of $\ell_{\text{spin}} \sim 23$ nm $\sim 80\, a$, indicating that the elementary excitations are highly mobile.

Figure 3A depicts the transverse thermal response along the $y$ axis, $\Delta T_y \equiv T_{L1} - T_{L2}$, at 8.3 K when the magnetic field is applied perpendicular to the 2D plane ($\boldsymbol{H} \parallel z$) in the presence of the thermal current along the $x$ axis. The longitudinal response due to misalignment of the contacts is cancelled by reversing the magnetic field. Asymmetric thermal response with respect to the field direction, $\Delta T_y^{\text{Asym}}(H) \equiv [\Delta T_y(+H) - \Delta T_y(-H)]/2$, is clearly resolved, establishing a finite thermal Hall effect. Special care was taken to detect the intrinsic thermal Hall signal from the sample (see Supplementary Information). Figure 3B depicts the temperature dependence of $-\kappa_{xy}/TB$ at the field of 15 T. Finite thermal Hall signal appears at $T \lesssim T^* \sim 60$ K. The sign of $\kappa_{xy}$ is negative, i.e. electron-like. As the temperature is lowered, $-\kappa_{xy}/TB$ first develops gradually and then increases steeply below 30 K. After reaching a maximum at around 15 K, $-\kappa_{xy}/TB$ decreases rapidly and changes sign slightly above $T_N$. It should be stressed that no discernible thermal Hall signal is observed in the paramagnetic state at $T \gtrsim T^* \sim 60$ K. Moreover, $-\kappa_{xy}/TB$ or $-\kappa_{xy}$ exhibits a peak at around $T_p$ where $\chi$ becomes maximum due to the development of the spin-spin correlation length (see the inset of Fig. 3B). These results lead us to conclude that the observed thermal Hall effect in volborthite arises from the magnetic excitations in the SL state, not from phonons (see Supplementary Information).

**Conclusions and Discussion.** The presence of gapless spin excitations revealed by $\chi$ and $C$ indicates that low-energy spin excitations in this insulating magnet have a similarity to fermions in paramagnetic metals with Fermi surface. Such a behavior has been predicted by the theory that features a spinon Fermi surface (6, 7, 8). The theory of Katsura, Nagaosa and Lee (17) emphasizes that deconfined fermionic spinons couple to an emergent gauge field. Through the coupling between the vector potential $\boldsymbol{A}$ and the gauge flux, the external magnetic field can exert a Lorentz force on spinons, which gives rise to a finite Hall effect, just as in the case of charged particles. Although applying this theory to the present system simply should be scrutinized, it is tempting to compare the Lorentz force acting on deconfined spinons with that acting on the free electrons. In a simple metal the thermal Hall angle $(\kappa_{xy}/T)/(\kappa_{xx}/T)$ equals to the electric Hall angle $\sigma_{xy}/\sigma_{xx}$ due to the Wiedemann-Franz law and is given as $(\kappa_{xy}/T)/(\kappa_{xx}/T) =$ $(eB)_e \tau_e / m_e$, where $m_e$ is the mass of electron, $\tau_e$ is the scattering time and $(eB)_e$ is the quantity proportional to the Lorentz force on the electron. By analogy, the Hall angle of deconfined spinons can be given as $(\kappa_{xy}/T)/(\kappa_{xx}^{\text{spin}}/T) = (eB)_{\text{spin}} \tau_{\text{spin}} / m_{\text{spin}}$. Here $m_{\text{spin}} \sim \hbar^2/(J_{eff} a^2) \sim 170 \, m_e$ and $\tau_{\text{spin}} = \ell_{\text{spin}}/v_{\text{spin}}$ is the effective mass and scattering time of spinons, respectively, and $(eB)_{\text{spin}}$ is proportional to the fictitious Lorentz force on the spinon. Using $\Delta\kappa/T \sim 0.03$ W K$^{-2}$ m$^{-1}$ and $\kappa_{xy} \sim -3 \times 10^{-4}$ W K$^{-1}$ m$^{-1}$ at 8 K, we obtain the upper limit of the Hall angle $(\kappa_{xy}/T)/(\kappa_{xx}^{\text{spin}}/T) \sim -1 \times 10^{-3}$ at 15 T. Thus the ratio of the Lorentz force acting on spinons and that on electrons is estimated to be $|(eB)_{\text{spin}}/(eB)_e| \sim 8 \times 10^{-3}$ at most. This implies that the coupling between the applied magnetic field and the gauge flux in a SL state is very small (29). Such a small coupling may be a reason of the absence of the quantum oscillations in a SL state of EtMe$_3$Sb[Pd(dmit)$_2$]$_2$ (ref. 3). We also note that this value is $\sim 4$ orders of magnitude smaller than the estimate reported in the paramagnetic state of Tb$_2$Ti$_2$O$_7$ (ref. 14).

At low temperatures below $T_p$ where the spin-spin correlation grows rapidly, $\kappa_{xy}/T$ shows a rapid reduction and changes the sign slightly above $T_N$. This suppression of $\kappa_{xy}/T \sim (C_{\text{spin}}/T)(\ell_{\text{spin}})^2 (eB)_{\text{spin}}/m_{\text{spin}}$ is unusual, because $C_{\text{spin}}/T$ and $\ell_{\text{spin}}$ do not show large temperature dependencies in this temperature range. There are several possible origins for this suppression. The suppression of $(eB)_{\text{spin}}$ may appear as a result of an instability of the fictitious gauge flux when the system approaches the ordered phase (29). Another possibility is the emergence of an additional spin excitations with the opposite sign of $(eB)_{\text{spin}}$. A kagomé structure gives rise to two types of hopping loops, triangular and hexagonal (17). The strong spin-spin correlations at low temperatures can generate spin excitations of the long hexagonal loop, which may induce a different sign of $\kappa_{xy}$. Further experimental studies on other frustrated SL systems may be important to clarify these nontrivial issues.

**Materials and Methods:**
The single crystals of volborthite used in this study were prepared by the hydrothermal synthesis. Typical sample size is $\sim 1$ mm $\times$ 0.5 mm $\times$ 0.07 mm. The crystals have an arrowhead-like shape with one twin boundary in the middle (26). All single crystals used in the thermal conductivity measurements were detwinned by cutting in half at the middle. The thermal conductivity and the thermal Hall conductivity were measured by the standard steady-state method in a variable temperature insert for 1.7 K to 80 K. An illustration of the setup is shown in the inset of Fig. 2 in the main text. A heat current ($Q$) was applied along the $b$ axis of volborthite, and magnetic field, $\mu_0 H \leq 15$ T, was applied perpendicular to the kagomé plane. We attached three Cernox$^{\text{TM}}$ thermometers (CX1050) and one heater on the sample. The configuration of the three thermometers ($T_{High}, T_{L1}, T_{L2}$) is shown in the inset of Fig. 2 in the main text. These thermometers were carefully calibrated in fields. We confirmed that all three thermometers show the identical magnetoresistance which matches that of the previous work (30) within our resolution.

The longitudinal thermal conductivity, $\kappa_{xx}$, and the thermal Hall conductivity, $\kappa_{xy}$, are obtained by

$$\frac{1}{wt}\begin{pmatrix} Q \\ 0 \end{pmatrix} = \begin{pmatrix} \kappa_{xx} & \kappa_{xy} \\ -\kappa_{xy} & \kappa_{xx} \end{pmatrix} \begin{pmatrix} (T_{High} - T_{L1})/L \\ (T_{L1} - T_{L2})/w \end{pmatrix} = \begin{pmatrix} \kappa_{xx} & \kappa_{xy} \\ -\kappa_{xy} & \kappa_{xx} \end{pmatrix} \begin{pmatrix} \Delta T_x/L \\ \Delta T_y/w \end{pmatrix}, \quad (S1)$$

where $\Delta T_x \equiv T_{High} - T_{L1}$, $\Delta T_y \equiv T_{L1} - T_{L2}$, $t$ is the thickness of the sample and $L$ ($w$) is the distance between the thermal contacts for $T_{High}$ and $T_{L1}$ ($T_{L1}$ and $T_{L2}$).

**Acknowledgments:**
We thank K. Behnia, H. Katsura, P.A. Lee, T. Momoi, O. I. Motrunich, N. Nagaosa, S.

Onoda, T. Senthil, A. Shitade, H. Takatsu, C. Varma for valuable discussions. This work was supported by Yamada Science Foundation, Toray Science Foundation and KAKENHI from JSPS and by a Grant-in-Aid for Scientific Research on Innovative Areas ``Topological Materials Science'' (KAKENHI Grant No. 15H05852).

**Author contributions:**

D.W., K.S, M.S, Y.S, H.I., T.S., Y.M. and M.Y. performed experiments and data analysis on single crystals prepared by T.Y., H.I. and Z.H. T.S., Y.M. and M.Y. wrote the paper. All authors critically reviewed the paper.

**Figure Legends**

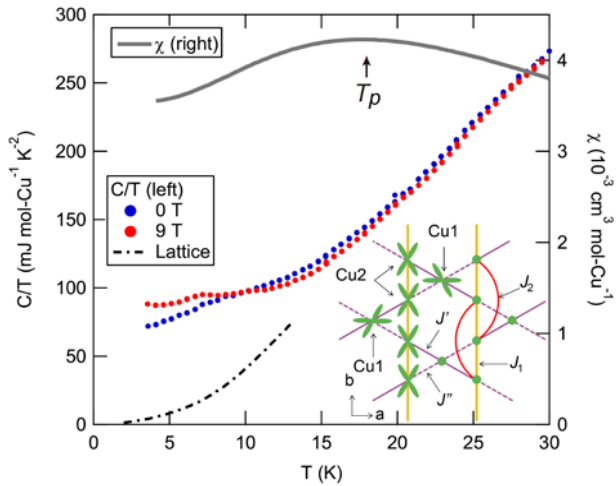

Figure 1: Temperature dependence of the heat capacity divided by temperature $C/T$ (closed circles, left axis) and the magnetic susceptibility $\chi$ (gray line, right axis) of a single crystal of volborthite. The peak temperature of the magnetic susceptibility is marked as $T_p$. Dashed line is the lattice heat capacity taken from ref. 31. At around 7 K, $C/T$ shows a kink in zero field, which is pronounced in the magnetic field of 9 T. However no signature of a magnetic transition has been observed in NMR and magnetic susceptibility measurements at this temperature. Therefore, the kink may be related to a lattice anomaly, though no discernible anomaly is observed in the thermal conductivity. The inset illustrates the arrangement of Cu ions in the *ab* plane. $J_1$ and $J_2$ represent the nearest-neighbor and next-nearest-neighbor interactions in the Cu2 spin chains, respectively. $J'$ and $J''$ represent the nearest-neighbor interactions between Cu1 and Cu2 spins.

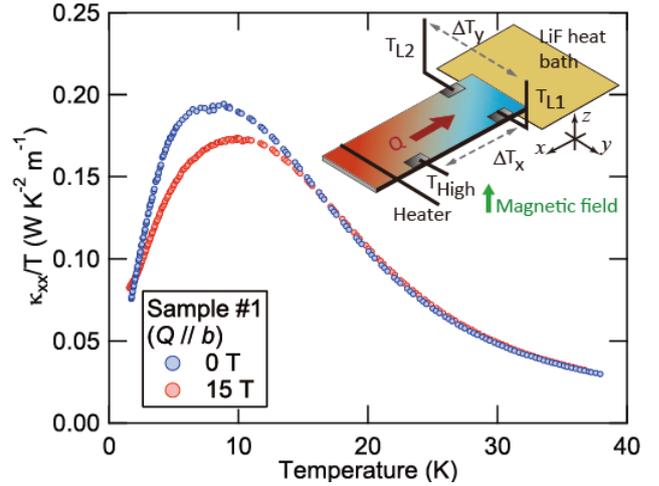

Figure 2: Temperature dependence of the longitudinal thermal conductivity divided by the temperature, $\kappa_{xx}/T$, at 0 (blue) and 15 T (red). The inset illustrates our experimental setup. Three thermometers ($T_{\text{high}}, T_{L1}, T_{L2}$) were attached to the sample. A heater was attached at one end of the sample to produce the thermal gradient along the *x* axis. The magnetic field was applied along the *z* axis. $\Delta T_x$ and $\Delta T_y$ were defined as $T_{\text{high}} - T_{L1}$ and $T_{L1} - T_{L2}$, respectively.

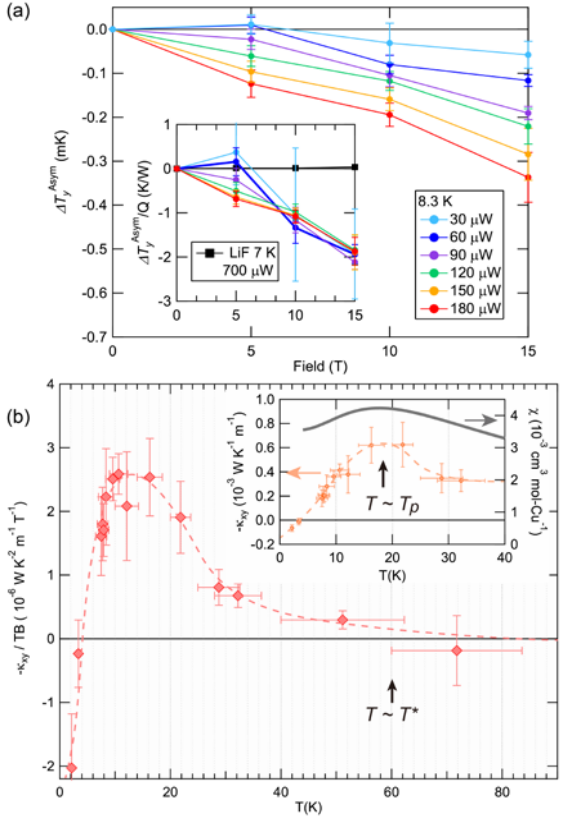

Figure 3: (a) The field and heater-power dependences of the antisymmetrized transverse temperature difference, $\Delta T_y^{\text{Asym}}(H) \equiv \left(\Delta T_y(+H) - \Delta T_y(-H)\right)/2$, at 8.3 K. The inset shows the same data normalized by the heater power, together with the data at 7 K of a LiF sample used as the heat bath (black squares). (b) Temperature dependence of $-\kappa_{xy}/TB$ at 15 T. Inset shows $-\kappa_{xy}$ at 15 T (left) and $\chi$ (right). The dotted lines are guides to the eyes. Error bars correspond to 1 s.d.

# Supporting Information

**Watanabe et al.**
**SI Text**

## S1. Tests to confirm the intrinsic thermal Hall signal

Since the antisymmetric component, $\Delta T_y^{\text{Asym}}(H) \equiv [\Delta T_y(+H) - \Delta T_y(-H)]/2$, is very small, we have made careful measurements and made several tests to confirm that the observed thermal Hall signal is intrinsic. First, to avoid thermal Hall signals from metals used in the cryostat, a LiF single crystal was used as the heat bath. Also, a non-metallic grease was used to attach the sample to the LiF heat bath. We checked the background signal of $\Delta T_y^{\text{Asym}}(H)$ from the LiF heat bath without the sample is negligibly smaller than that of volborthite (see the inset of Fig. 3A) in all temperature and field ranges of this study. Second, dependences of $\Delta T_y^{\text{Asym}}(H)$ on the heater power were checked at the same temperature (main panel of Fig. 3A). As shown in the inset of Fig. 3A, $\Delta T_y^{\text{Asym}}(H)$ show a linear response to the heater power within our resolution. Finally, we reversed the thermal contacts to the thermometers L1 and L2 with keeping other setups, and confirmed that the antisymmetrized temperature gradient along the $y$ axis shows the same sign (not shown, the antisymmetrized $T_{L1} - T_{L2}$ shows the reversed sign). These tests can exclude the possibility that the observed signal is an artifact, such as errors of field calibration of the thermometers or thermal leaks from the heater to the thermometers. Hence, we can safely confirm the observed $\Delta T_y^{\text{Asym}}(H)$ as intrinsic signals.

## S2. Thermal Hall effect of phonons

Here, we note that the phonon Hall effect is unlikely as the origin of $\kappa_{xy}$ of volborthite. The thermal Hall effect of phonons has been reported in dielectric garnet $Tb_3Ga_5O_{12}$ (TbGG) (32). Theoretically, skew-like scattering effects of phonons by large spins (33-36) have been discussed as the origin of the phonon Hall effect. In fact, a large magneto-elastic coupling of TbGG has been shown by acoustic measurements (37, 38). The strong scattering effects of phonons by spins has also been shown by the small thermal conductivity of TbGG compared by other rare-earth gallium garnets (39). The complex crystal-field structure of quasi-doublets of Tb ions with a large magnetic moment (40) has been pointed out as the origin of the strong spin-phonon coupling (37, 38). Obviously, these situations are completely different from that of volborthite, where spin-1/2 $Cu^{2+}$ ions form a frustrated antiferromagnet. Moreover, $\kappa_{xy}$ of phonons should show a larger signal at a higher temperature where the phonon contribution becomes dominant. In contrast, we find the $\kappa_{xy}$ decreases as the temperature increases above 20 K where the field dependence of $\kappa_{xx}$ (Fig. S1) shows a dominant contribution of $\kappa_{xx}^{\text{ph}}$. We, therefore, conclude that the thermal Hall signal is due to the Hall effect on the spin excitations.

**Supplementary Figure Legends**

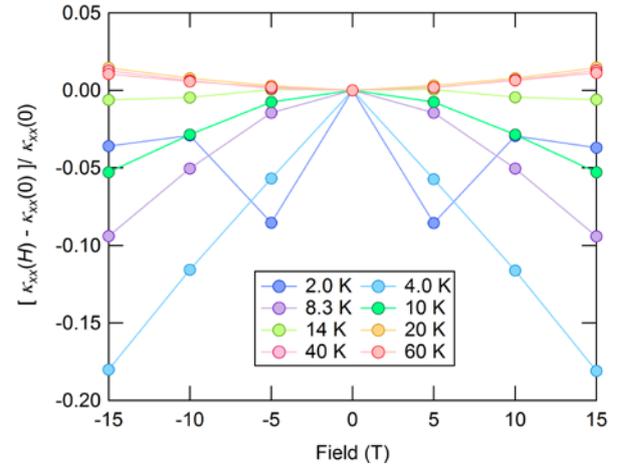

**Fig. S1** The field dependence of $\kappa_{xx}$ at different temperatures. The normalized difference of $\kappa_{xx}$ under field, $\Delta\kappa_{xx}(H)/\kappa_{xx}(0) \equiv (\kappa_{xx}(H) - \kappa_{xx}(0))/\kappa_{xx}(0)$, is plotted as a function of field. At the lowest temperature of our measurement (2.0 K), $\kappa_{xx}$ increases above $\sim 5$ T, which is due to a magnon contribution in the SDW phase (25). At higher temperature above $\sim 20$ K, $\kappa_{xx}$ turns to increase under fields due to the increase of $\kappa_{xx}^{\text{ph}}$ by applying fields.